\documentclass[aps,prc,twocolumn,showpacs,floatfix,nofootinbib]{revtex4}
\usepackage{amsmath,graphicx}
\newcommand{\avgev}[1]{\left\langle{#1}\right\rangle}
\begin{document}

\title{Effects of initial-state dynamics on collective flow within a
  coupled transport and viscous hydrodynamic approach}

\author{Chandrodoy Chattopadhyay$^1$, Rajeev S. Bhalerao$^2$,
  Jean-Yves Ollitrault$^3$, and Subrata Pal$^1$}

\affiliation{$^1$Department of Nuclear and Atomic Physics, Tata
  Institute of Fundamental Research, Homi Bhabha Road, Mumbai 400005,
  India}

\affiliation{$^2$Department of Physics, Indian Institute of Science
  Education and Research (IISER), Homi Bhabha Road, Pune 411008,
  India}

\affiliation{$^3$CNRS, URA2306, IPhT, Institut de physique th\'eorique
  de Saclay, F-91191 Gif-sur-Yvette, France}

\begin{abstract}
We evaluate the effects of preequilibrium dynamics on observables in 
ultrarelativistic heavy-ion collisions. 
We simulate the initial nonequilibrium phase within A MultiPhase Transport
(AMPT) model, 
while the subsequent near-equilibrium evolution 
is modeled using (2+1)-dimensional relativistic viscous hydrodynamics. 
We match the two stages of evolution carefully by calculating the
full energy-momentum tensor from AMPT and using it as input for the
hydrodynamic evolution. 
We find that when the preequilibrium evolution is taken into account,
final-state observables are insensitive to the switching time from
AMPT to hydrodynamics. 
Unlike some earlier treatments of preequilibrium dynamics, we do not find the initial shear viscous tensor to be large.
With a shear viscosity to entropy density ratio of $0.12$, our model
describes quantitatively a large set of experimental data on Pb+Pb
collisions at the Large Hadron Collider(LHC) over a wide range of centrality: 
differential anisotropic flow $v_n(p_T) ~(n=2-6)$,  
event-plane correlations, 
correlation between $v_2$ and $v_3$, and 
cumulant ratio $v_2\{4\}/v_2\{2\}$. 
\end{abstract}



\maketitle

\section{Introduction}

High-energy heavy-ion collision experiments at the Relativistic
Heavy-Ion Collider (RHIC) \cite{Adams:2005dq,Adcox:2004mh} and at the
Large Hadron Collider (LHC)
\cite{ALICE:2011ab,ATLAS:2012at,Chatrchyan:2013kba} have established
the formation of a strongly-interacting Quark-Gluon Plasma (QGP).
Evidence is based on the large collective flow observed in the plane
transverse to the beam axis, in particular the anisotropic flow. 
 These observations can be explained by treating the formed QGP as a viscous relativistic fluid 
\cite{Israel:1979wp,Muronga:2003ta,Romatschke:2007mq,Song:2007ux},
with a small shear viscosity to entropy density ratio 
$\eta/s$~\cite{Heinz:2013th}, corresponding to a strongly-interacting
system~\cite{Kovtun:2004de}. 
The flow is found to originate mostly from the early, partonic stage
of the expansion. 
It is therefore essential to scrutinize its sensitivity to the early
dynamics, in particular, to the early stages where hydrodynamics
cannot be applied.

The initial stage, defined as the stage after which the hydrodynamic description
is permissible, is the largest source of uncertainty in hydrodynamic
modeling. 
Not only is the initial energy density profile poorly constrained~\cite{Retinskaya:2013gca,Moreland:2014oya}, 
the matter formed is also out of equilibrium in several respects. 
First, the expansion into the vacuum generates significant transverse
flow at early times, which must be taken into account when setting up 
realistic initial conditions for 
hydrodynamics~\cite{Vredevoogd:2008id,Liu:2015nwa}. 
Second, due to the rapid longitudinal expansion, the pressure is
strongly anisotropic at early times~\cite{Gelis:2013rba} (the
longitudinal pressure is smaller than the transverse pressure) 
which has triggered the development of ``anisotropic
hydrodynamics''~\cite{Florkowski:2010cf,Martinez:2010sc}.
Both effects, initial flow and pressure anisotropy, are encoded
in the energy-momentum tensor $T^{\mu\nu}$ used as an initial
condition for hydrodynamic calculations. 
Therefore, a proper approach to preequilibrium dynamics is to model the 
full $T^{\mu\nu}$. 
This has first been done in the context of
strong-coupling calculations~\cite{vanderSchee:2013pia}, and more
recently in the weak-coupling regime~\cite{Keegan:2016cpi,Kurkela:2017hgm}. 
However, there are to date few hydrodynamic calculations using as
input the full energy-momentum tensor 
$T^{\mu\nu}$ resulting from a consistent model of the early
dynamics~\cite{vanderSchee:2013pia,Mantysaari:2017cni}.  
For instance, the
  IP-Glasma+MUSIC calculation of Ref.~\cite{Gale:2012rq} does not
  conserve the full $T^{\mu\nu}$ when switching from classical gluon
  dynamics to hydrodynamics and neither does the recent 
superSONIC calculation of Ref.~\cite{Weller:2017tsr}. 

In this article, we use the multiphase transport model AMPT
\cite{Lin:2004en} to model the preequilibrium dynamics. 
AMPT implements realistic cross sections between particles. 
It thus complements previous idealized approaches using weak coupling
or strong coupling techniques. 
AMPT is able to simulate the entire collision event, but we use it 
here only to model the initial stages.  
It has been used earlier as an input to 
ideal~\cite{Pang:2012he} and 
viscous~\cite{Bhalerao:2015iya,Zhao:2017yhj} hydrodynamic
calculations, but at the expense of 
discarding part of the information contained in $T^{\mu\nu}$. 
Here, we switch from AMPT to (2+1)-dimensional second-order viscous
hydrodynamics~\cite{Shen:2014vra} by matching the full $T^{\mu\nu}$. 
The details of this hybrid model are described in Sec.~\ref{s:model}. 
In Sec.~\ref{s:preequilibrium}, we discuss the sensitivity of
hydrodynamic flow to the initial stages. 
In Sec.~\ref{s:data}, we compare the results of our model 
with several LHC data on Pb+Pb collisions at 2.76 and 5.02 TeV: 
transverse-momentum spectra, anisotropic flow, correlations between
flow magnitudes in different harmonics, two- and three-event-plane
correlators.

\section{The model and the initial conditions}
\label{s:model}

The AMPT model~\cite{Lin:2004en} is a widely-used transport model
which provides a good description of several observables of heavy-ion
collisions, 
in particular pair correlations~\cite{Adam:2016tsv} and 
anisotropic flow~\cite{Xu:2011jm,Pal:2012tc}, over
a wide range of colliding energies~\cite{Adamczyk:2013gw}. 
AMPT has also been able to predict quantitatively the magnitudes of event-plane 
correlations~\cite{Bhalerao:2013ina,Aad:2014fla} and other 
multiparticle correlations~\cite{Acharya:2017gsw}. 
The AMPT version implemented in this paper uses 
the HIJING 2.0 model~\cite{Deng:2010mv,Pal:2012gf} to 
determine the nucleon configuration in an event. 
Nucleons can undergo soft collisions, which lead to string 
excitations, and hard collisions, which produce minijet 
partons~\cite{Wang:1991hta}.  
We have 
employed the string melting version of AMPT \cite{Lin:2004en}, 
in which strings are melted into their constituent quarks and
antiquarks, and which improves the description of the anisotropic flow data. 
The scatterings among these quarks and minijet partons
and their evolution are treated with ZPC parton
cascade~\cite{Zhang:1997ej} with a parton-parton elastic cross section
of 1.5~mb.   

While AMPT by itself can simulate the entire collision event, 
we use it here only to describe the first stages, and then couple it
to a viscous hydrodynamic description. 
The hydrodynamic code we use~\cite{Shen:2014vra} is 2+1 dimensional, in
the sense that it 
assumes boost invariance in the longitudinal
direction~\cite{Bjorken:1982qr} and determines numerically the
transverse flow only. 
This choice is motivated by simplicity, and by the observation that
anisotropic flow depends little on 
rapidity~\cite{Alver:2010rt,Chatrchyan:2012wg,ATLAS:2012at} 
We thereby neglect the effect of longitudinal
fluctuations~\cite{Bozek:2010vz,Petersen:2011fp,Xiao:2012uw,Jia:2014ysa,Pang:2014pxa,Pang:2015zrq,Khachatryan:2015oea,Aaboud:2017tql,Bozek:2017qir},  
which have been much studied lately, but mildly affect flow  
observables near midrapidity. 
The transition from AMPT to hydrodynamics is implemented on a constant
proper time hypersurface $\sqrt{t^2-z^2}=\tau_{\rm sw}$. 
Since the AMPT model is 3+1 dimensional, we need to project it as we
switch to the 2+1 dimensional hydrodynamic model. 
This is achieved by averaging over the space-time
rapidity $\eta_s$, 
defined as $\eta_s \equiv(1/2)\ln[(t+z)/(t-z)]$, 
in the window $-3 < \eta_s< 3$.\footnote{We choose a large rapidity
  window to maximize the statistics.} 
We include all particles in this window and consider only
their longitudinal momenta relative to the fluid. 
In the Bjorken picture~\cite{Bjorken:1982qr}, the longitudinal fluid
velocity is $v_z=z/t$. 
Therefore, the longitudinal motion relative to the fluid is obtained
by transforming the energy and longitudinal momentum as follows:
\begin{eqnarray}
E' &=& E \cosh\eta_s - p_z \sinh\eta_s, \nonumber\\
p_z' &=& p_z \cosh\eta_s - E \sinh \eta_s.
\end{eqnarray}
The transverse momentum is unchanged: $p_T' = p_T$. 

We now describe how the energy-momentum $T^{\mu\nu}$ is evaluated. 
Switching from a discrete description, in terms of pointlike
particles, to a continuous description, in terms of a fluid, typically
involves a coarse-graining procedure, where one defines a fluid
element by the particles it contains. 
We choose an alternative procedure and treat each particle as an 
extended object, whose size is much larger than the transverse
distance between particles, so that the fluid formed by all the
particles is smooth.  
Specifically, we smear each parton in AMPT by 2D
Gaussian distribution in the transverse plane~\cite{Steinheimer:2007iy}. 
$T^{\mu\nu}$ is defined at each point as 
\begin{eqnarray}\label{smear}
T^{\mu\nu}(x,y) &=& \frac{1}{2\pi\sigma^2\tau_{\rm sw}\Delta\eta_s} 
\sum_i \frac{p_i'^\mu p_i'^\nu}{p_i'^0} \nonumber \\  
&& \times \exp \left[ - \frac{(x-x_i)^2 + (y-y_i)^2}{2\sigma^2} \right] ,
\end{eqnarray}
where the sum runs over all partons $i$ with transverse coordinates
($x_i, ~y_i$) and energies $E_i' \equiv p_i'^0 = \sqrt{{\bf p}_i'^2 +
  m_i^2}$, and 
$\Delta\eta_s=6$ is the width of the $\eta_s$ window. 
The Gaussian transverse width is a free parameter which we set to $\sigma =0.8$~fm. 

The energy-momentum tensor in viscous hydrodynamics is usually written
as~\cite{Romatschke:2009im}
\begin{equation}\label{NTD}
T^{\mu\nu} = \epsilon u^\mu u^\nu - (P+\Pi) \Delta^{\mu \nu} +
\pi^{\mu\nu}, 
\end{equation}
where $u^\mu$ is the fluid 4-velocity, $\epsilon$ and $P$ are the
energy density and pressure in the fluid's local rest frame, 
$\Delta^{\mu\nu}=g^{\mu\nu}-u^\mu u^\nu$ is the projection operator on
the three-space orthogonal to $u^\mu$ defined in the Landau frame, 
$\Pi$ is the bulk pressure, and $\pi^{\mu\nu}$ is the shear pressure
tensor. 
We now explain how the quantities in the right-hand side of
Eq.~(\ref{NTD}) are obtained from $T^{\mu\nu}$. 
$\epsilon$ and $u^\mu$ are given by the Landau matching condition:
\begin{equation}\label{eigen}
T^{\mu\nu} u_\nu = \epsilon u^\mu . 
\end{equation}
The pressure $P$ is then related to $\epsilon$ by the equation of
state. 
We have employed the s95p-PCE
equation of state~\cite{Huovinen:2009yb} which is obtained from fits
to lattice data for crossover transition and matches a realistic
hadron resonance gas model at low temperatures $T$, with partial
chemical equilibrium (PCE) of the hadrons at temperatures below
$T_{\rm PCE} \approx 165$ MeV.

The bulk pressure $\Pi$ is then obtained from the trace:
\begin{equation}\label{bulk}
T^{\mu}_{\mu}=\epsilon-3(P+\Pi).
\end{equation}
Using Eq.~(\ref{smear}), the contribution of each parton to
$T^{\mu}_{\mu}$ is proportional to $p^\mu p_\mu=m^2$. 
The masses of partons in AMPT are current quark masses, which are
small for light quarks, 
so that the bulk pressure $\Pi$ is small. We neglect it in 
the present calculation.  
Finally, the shear pressure tensor $\pi^{\mu\nu}$ is 
given by Eq. (\ref{NTD}). 

Our procedure conserves the full structure of the energy momentum
tensor from the initial stage, and therefore automatically includes
the effect of initial transverse flow and a viscous corrections to the
pressure tensor. 
The resulting initial conditions are more realistic than 
typical prescriptions where $\pi^{\mu\nu}$ is set to
0~\cite{Weller:2017tsr} or  initialized 
to the Navier-Stokes value~\cite{Luzum:2008cw}.
Our transport calculation also takes into account interactions 
before the start of hydrodynamics. The resulting hybrid 
calculation is more consistent in this respect than that of
Ref.~\cite{Liu:2015nwa}, where the pre-equilibrium stage is modeled by
free-streaming partons.  
Naturally, at the instant of switch-over to hydrodynamics, the system
in  Ref.~\cite{Liu:2015nwa} is far from equilibrium with a large shear
viscous tensor, whereas in the present work the preequilibrium
dynamics drives the system close to local equilibrium, allowing a
smooth matching to the hydrodynamics at the switch-over time. 
We account for the full preequilibrium dynamics, as was done
previously in Refs.~\cite{vanderSchee:2013pia,Mantysaari:2017cni}.  

The hydrodynamic evolution is continued till each fluid cell reaches a 
decoupling temperature of $T_{\rm dec} =120$ MeV. The hadronic spectra
are obtained at this temperature using the Cooper-Frye
prescription~\cite{Cooper:1974mv}:
\begin{equation}\label{CF}
\frac{dN}{d^2p_Tdy} = \frac{g}{(2\pi)^3} \int p_\mu d\Sigma^\mu f(x,p),
\end{equation}
where $g$ is the degeneracy, 
$p^\mu$ is the four-momentum of the particle,
$d\Sigma^\mu$ represents the element of the 3D freeze-out
hypersurface and $f(x,p) = f_0 + \delta f$ is the nonequilibrium
phase-space distribution function at freezeout.  
We have used the standard viscous correction form corresponding to
Grad's 14-moment approximation~\cite{Teaney:2003kp}:
\begin{equation}\label{Grad}
\delta f = \frac{f_0 \tilde f_0}{2(\epsilon+P)T^2}\, p^\alpha p^\beta \pi_{\alpha\beta},
\end{equation}
where corrections up to second order in momenta are present, and
$\tilde f_0 \equiv 1-r f_0$, with $r=1,-1,0$, are the equilibrium
distributions for the Fermi, Bose, and Boltzmann gases, respectively.
Resonances of masses up to about 2.25 GeV are included in the
calculations to be consistent with the s95p-PCE equation of state, and
the results presented include the resonance decays.

In this work, we neglect the temperature dependence of the shear
viscosity over entropy ratio $\eta/s$~\cite{Niemi:2015qia}. 
We choose the value $\eta/s=0.12$ which gives a good description of
anisotropic flow data (see  Sec.~\ref{s:data}).

The initial conditions from AMPT need to be  adjusted.  The reason is that 
the multiplicity obtained in the AMPT+hydrodynamics model 
is slightly smaller than that obtained using the AMPT model 
alone, which matches experimental data.  
This can be due to the projection from 3 to 2 dimensions when we
switch from AMPT to hydrodynamics, or from a difference between the
effective viscosity in the AMPT calculation and that
used in the hydrodynamic calculation. 
We therefore rescale the initial energy density profile of the 
hydrodynamic calculation~\cite{Liu:2015nwa} by a constant factor. 
The value of this factor, which is roughly $1.2$, is determined 
by fitting the charged multiplicity density $dN_{\rm   ch}/dy$ to
$2.76$~TeV LHC data in the $0-5\%$ centrality 
range. We use the same factor for other centralities and other
colliding energies.

\section{Effects of preequilibrium dynamics} 
\label{s:preequilibrium}

We study the sensitivity of collective flow 
to the preequilibrium dynamics. 
For this purpose, we generate 300 Pb+Pb collisions at
$\sqrt{s_{NN}}=5.02$~TeV in the 40-50\% centrality interval, 
where elliptic flow in the reaction plane is largest~\cite{Aamodt:2010pa}. 
Throughout this article, the centrality $c$ is defined according to impact
parameter $b$ by $c=\pi b^2/\sigma$~\cite{Broniowski:2001ei},
with a nucleus-nucleus total inelastic cross-section of $\sigma = 784$ and
796~fm$^2$ for collisions at $\sqrt{s_{NN}}=2.76$ and 5.02 TeV,
respectively, as calculated from the Glauber model~\cite{Bhalerao:2015iya}. 

\begin{figure}[t]
\begin{center}
\scalebox{0.32}{\includegraphics{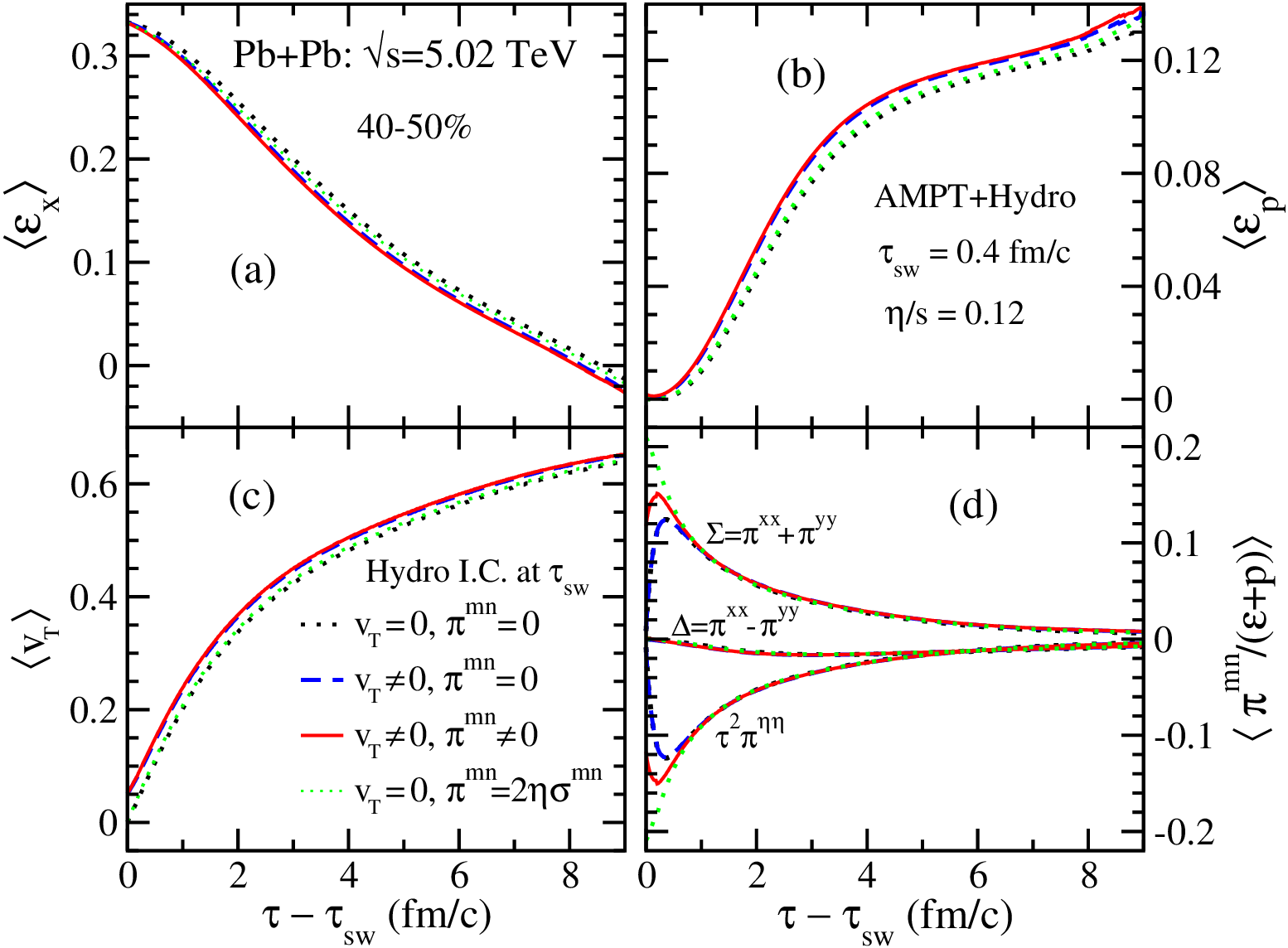}}
\end{center}
\vspace{-0.4cm}
\caption{(Color online) Time evolution of the eccentricity in the
  reaction plane (a), the momentum anisotropy (b), 
the transverse flow velocity (c) and various components of the
shear pressure tensor $\pi^{mn}$ normalized by the enthalpy density
(d). 
Averages over the transverse plane in panels (a), (c) and (d) are 
evaluated with a Lorentz contracted energy density as
  weight~\cite{Liu:2015nwa}. All quantities are averaged over events. 
Each panel compares four different initializations (see text). 
The switching time between AMPT and hydrodynamics is $\tau_{\rm
  sw}=0.4$ fm/$c$.} 
\label{fig:timevol}
\end{figure}

\begin{figure}[t]
\begin{center}
\scalebox{0.37}{\includegraphics{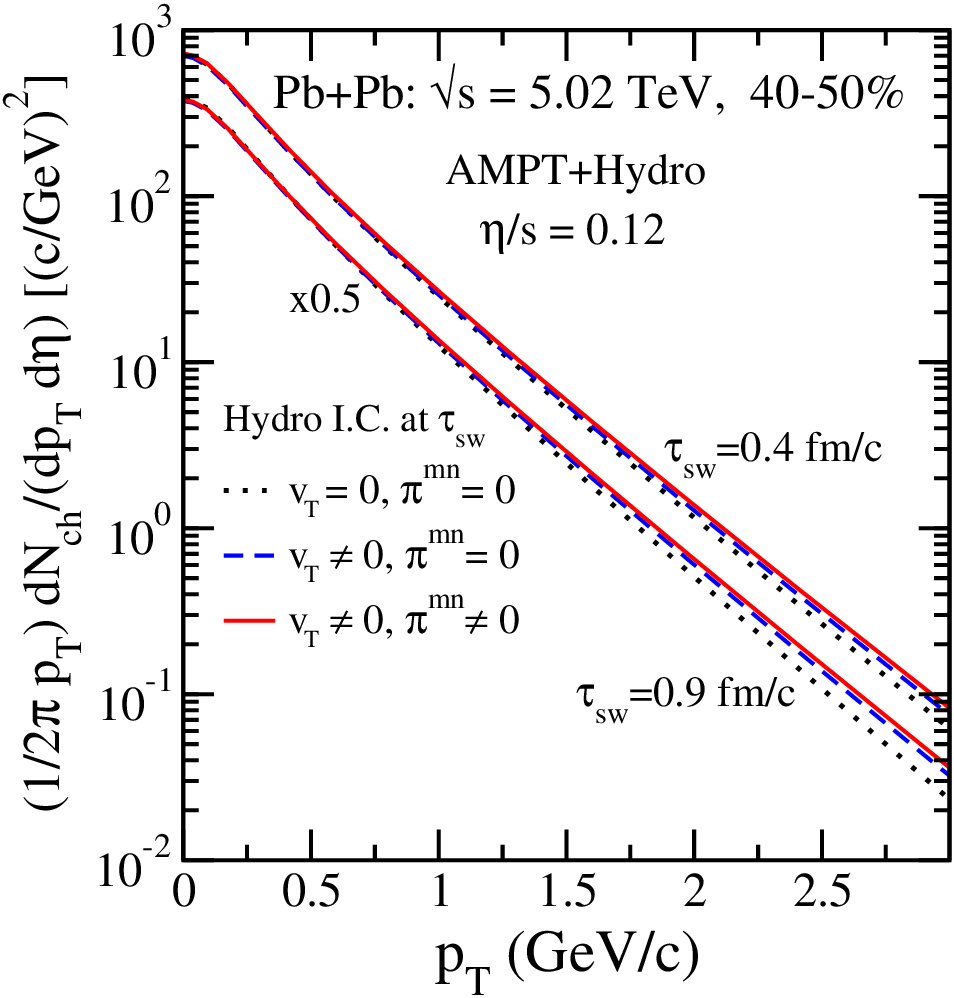}}
\end{center}
\vspace{-0.4cm}
\caption{(Color online) Transverse momentum spectra of charged
  particles from the AMPT+Hydro calculations. 
The results are
  for two switching times $\tau_{\rm sw}=0.4$ and 0.9 fm/$c$ and three
  different initial conditions. Curves for $\tau_{\rm
    sw}=0.9$~fm/c are shifted vertically in order to avoid
  overlapping. 
}
\label{fig:dndpt}
\end{figure}

We first test the sensitivity to the preequilibrium dynamics by varying the
initialization of the hydrodynamic calculation. 
Specifically, we compare the default version, where the full 
$T^{\mu\nu}$ is evaluated using 
AMPT (shown as red solid lines in Figs.~\ref{fig:timevol} and
\ref{fig:dndpt}), with three simplified versions:
a fully simplified version where both the initial shear tensor
$\pi^{\mu\nu}$ and the transverse velocity 
$v_T$ are set to 0 (black dotted lines), corresponding to traditional hydrodynamic
calculations~\cite{Luzum:2008cw} where the initial conditions are
specified solely by the initial energy density profile; 
one where the transverse velocity is set to zero and the shear tensor
to the Navier-Stokes value
$\pi^{\mu\nu}=2\eta\sigma^{\mu\nu}$~\cite{Song:2007ux,Luzum:2008cw}
(green dotted lines); 
and one where one keeps the transverse velocity but sets 
the shear tensor $\pi^{\mu\nu}$ to 0 (blue dashed lines). 
In this way, we can test separately the effects of initial flow and 
initial shear tensor.

Figure~\ref{fig:timevol} displays the time evolution of various
quantities in the hydrodynamic phase. The spatial eccentricity 
 in the reaction plane
$\varepsilon_x$~\cite{Voloshin:1999gs,Teaney:2010vd} 
  is shown in panel (a).  
Its initial value is large, corresponding to the almond-shaped area of
the overlap region between the nuclei (the impact parameter is in the
range $10-11$~fm). 
As the system expands in all directions, its shape becomes rounder and
the spatial eccentricity decreases~\cite{Kolb:2000sd}. 
This decrease is slightly faster if initial transverse flow is
included. 

The spatial eccentricity creates a momentum anisotropy due to pressure
gradients~\cite{Ollitrault:1992bk}, corresponding to elliptic flow. The momentum anisotropy
is defined as~\cite{Luzum:2008cw}
\begin{equation}\label{ellip}
\varepsilon_p \equiv \frac{\int d^2r_\perp (T^{xx} - T^{yy})} 
{\int d^2r_\perp (T^{xx} + T^{yy})}.
\end{equation}
Figure~\ref{fig:timevol} (b) shows that $\varepsilon_p$ develops 
in the first few fm/c of the expansion~\cite{Sorge:1996pc}, as the
spatial anisotropy $\varepsilon_x$ decreases. 
The sensitivity of  $\varepsilon_p$  to preequilibrium dynamics is
small, but clearly visible.\footnote{We switch to hydrodynamics at an
  early time  $\tau_{\rm   sw}=0.4$ fm/$c$, therefore the
  preequilibrium phase does not last long and its effect is limited.}
When initial transverse flow is present, the momentum anisotropy is
larger and develops earlier. 
Note, however, that the value of $\varepsilon_p$ at $\tau=\tau_{\rm sw}$
is close to 0, even if initial flow is included. 
The effect of the initial shear pressure is much smaller than that of
initial flow. It only increases slightly the anisotropy, due to the
larger transverse pressure.  
The mean transverse flow velocity, displayed in panel (c), follows the
same pattern, showing that radial flow and elliptic flow are closely
related. 
The inclusion of initial flow imparts a 
small transverse kick (about 5\% of the speed of light) at $\tau_{\rm
  sw}$. 

Figure \ref{fig:timevol}(d) shows the time evolution of the dominant
components of the viscous pressure tensor, namely,
$\tau^2\pi^{\eta\eta}$, the sum $\Sigma = \pi^{xx} + \pi^{yy}$ and the
difference $\Delta = \pi^{xx} - \pi^{yy}$, all normalized by the
enthalpy density. The initial values arising from the AMPT
preequilibrium dynamics are about 12\% for the first two of these
three components. This is in contrast with Ref.~\cite{Gale:2012rq}
where $\pi^{\mu\nu}$ arising from the preequilibrium dynamics was too
large and had to be arbitrarily set to zero.\footnote{This reference, however,
does not give quantitative information about the viscous tensor. 
Since we have a more realistic microscopic dynamics in the early stage
including  realistic parton-parton elastic 
cross sections, it is plausible that our early dynamics drives the
system toward local equilibrium more efficiently.}
Note also that AMPT initial values for $\tau_{\rm sw}^2\pi^{\eta\eta}$
and $\Sigma$ are about $40\%$ smaller than Navier-Stokes values. 
As time evolves, the
magnitudes of all three components at first increase due to additional
contributions from the viscous hydrodynamics VISH2+1. Thereafter, they
all decrease and become negligible at late times
\cite{Song:2007ux}. 
The sensitivity to the initial value of $\pi^{\mu\nu}(\tau_{\rm sw})$
(0, Navier-Stokes, or AMPT) is only visible in the first 1~fm/c: 
The curves then all converge to the same value. 
This explains why the results shown in panels (b) and (c) have little
sensitivity to the initial value of the shear tensor. 

Figure \ref{fig:dndpt} shows the transverse momentum spectra of
charged hadrons with different initialization schemes. 
Transverse flow tends to increase the transverse momentum. 
We have seen in Fig.~\ref{fig:timevol} (c) that initial flow increases
the transverse flow at later times. Therefore, it results in more particles at larger
$p_T$. Inclusion of the initial shear tensor has a smaller effect, and
goes in the same direction. 
We also display results with a larger value of the switching time 
$\tau_{\rm sw} =0.9$~fm/$c$. 
This leads to a larger initial flow from preequilibrium dynamics in AMPT
but leaves less time to develop hydrodynamic flow in VISH2+1. The net 
effect is a slightly softer spectrum compared to that for 0.4 fm/$c$.

\begin{figure*}[t]
\begin{center}
\scalebox{0.40}{\includegraphics{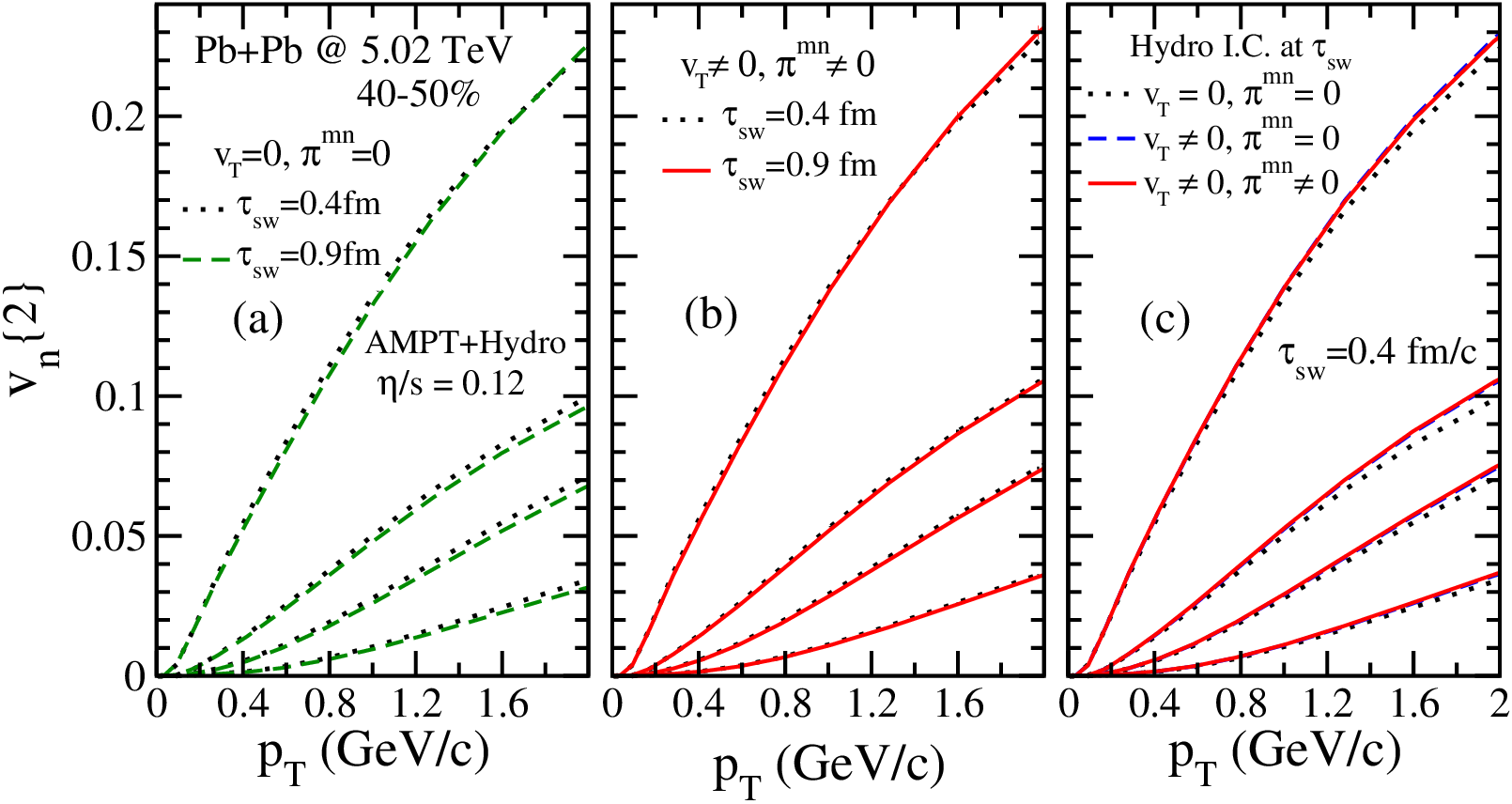}}
\end{center}
\vspace{-0.4cm}
\caption{(Color online) 
  Anisotropic flow coefficients, $v_n\{2\}(p_T)$ 
  (for $n=2-5$, top to
  bottom), for charged hadrons in the AMPT+Hydro calculations.
(a) Initial flow and viscous tensor set to 0, and two different
  switching times, $\tau_{\rm  sw}=0.4$ fm/$c$ (black dotted lines)
  and 0.9 fm/$c$ (green dashed lines). 
(b) With initial flow and viscous tensor from AMPT. 
(c) Results for three initial conditions at $\tau_{\rm sw}=0.4$ fm/$c$.}
 \label{vnpt_t49_b45_LHC502}
\end{figure*}

Figure~\ref{vnpt_t49_b45_LHC502} displays the anisotropic flow
coefficients $v_2(p_T)$ to $v_5(p_T)$. 
They are computed for each hydro event using the usual
formulas~\cite{Heinz:2013bua}. The average over events is evaluated in a way 
that closely follows the experimental procedure: 
 $v_n(p_T)$ is measured by correlating a particle in
a given $p_T$ window with a second particle belonging to the same event, but
without any restriction on $p_T$, and then averaging over events. 
The corresponding formulas in hydrodynamics are written explicitly in 
Ref.~\cite{Heinz:2013bua}. Specifically, the quantity we evaluate is the 
``two-particle cumulant flow'' as defined in this reference.  

Figure \ref{vnpt_t49_b45_LHC502}(a) compares the values of $v_n$
obtained for two different switching times $\tau_{\rm sw}=0.4$ fm/$c$
(black dotted lines) and 0.9 fm/$c$ (green dashed lines). 
In this
calculation both the transverse velocity and viscous tensor are set to
zero at the switching times. 
Hence, any preequilibrium build-up of flow is ignored here. 
A delayed start of hydrodynamics at $\tau_{\rm
  sw}=0.9$ fm/$c$ leaves less time for the hydrodynamic 
build-up of momentum anisotropy. 
This causes a slight reduction in $v_n(p_T)$ as compared to
the earlier switching time 0.4 fm/$c$. The effect is more pronounced
for higher flow harmonics. 
Figure
\ref{vnpt_t49_b45_LHC502}(b) is similar to
\ref{vnpt_t49_b45_LHC502}(a) except that the full preequilibrium
dynamics from AMPT is included. 
This results in a slight increase of $v_n$, as expected from
Fig.~\ref{fig:timevol} (b). 
Remarkably, the sensitivity to the switching time becomes negligible
once preequilibrium dynamics is included. 
This means that it is essentially equivalent to run AMPT or
viscous hydrodynamics at early times~\cite{Vredevoogd:2008id}. 

In Fig. \ref{vnpt_t49_b45_LHC502}(c) we compare $v_n(p_T)$ for various
initial conditions at a fixed $\tau_{\rm sw}=0.4$ fm/$c$. Compared to
the initial $v_T=0$ case, the inclusion of transverse flow at the
switching time injects an additional (finite but small) flow
anisotropy at the start of VISH2+1 (see Fig. \ref{fig:timevol}(b)). This
results in a slight enhancement of $v_n(p_T)$ for nonzero flow
initialization (blue dashed lines) as compared to the flow-free case
(black dotted lines). Further inclusion of viscous tensor has
insignificant effect on $v_n(p_T)$ (red solid lines) as hydrodynamic
evolution ceases to remember the initial $\pi^{\mu\nu}$ values 
(see Fig. \ref{fig:timevol}(d)).

In summary, preequilibrium dynamics increases the transverse flow, but
this is a small increase as long as the switching time is small. The
calculations presented in the next section are carried out with the
full preequilibrium dynamics from AMPT. 

\section{Comparison with LHC data}
\label{s:data}

\begin{figure}[b]
\begin{center}
\scalebox{0.35}{\includegraphics{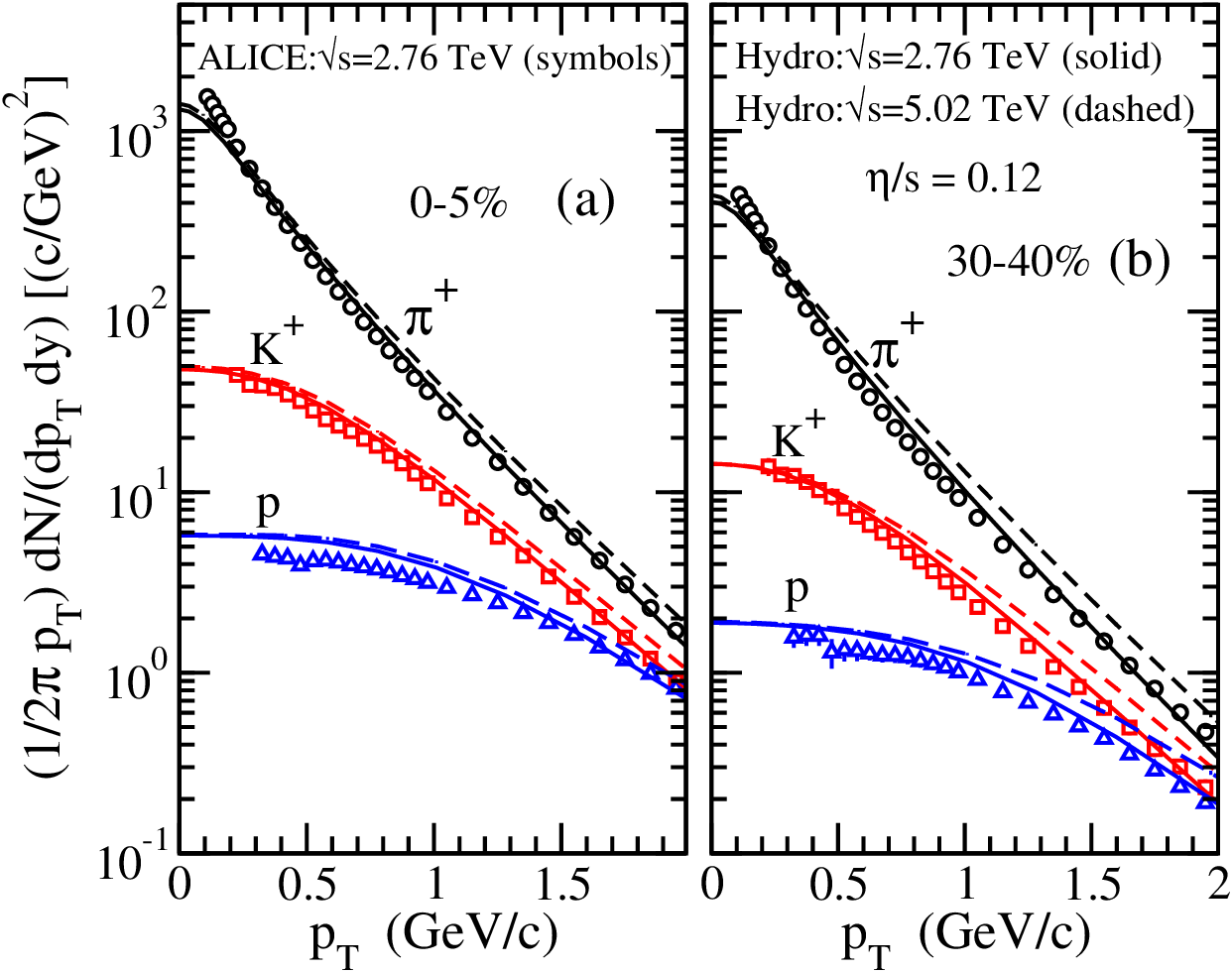}}
\end{center}
\vspace{-0.4cm}
\caption{(Color online) Transverse momentum spectra of pions, kaons,
  and protons at midrapidity for two centrality ranges, $0-5$\% and
  $30-40$\% in Pb+Pb collisions at $\sqrt{s_{NN}}=2.76$ TeV in the
  AMPT+Hydro model (solid lines) as compared to the ALICE data
  \cite{Abelev:2013vea} (symbols). Model predictions of
  the particle spectra for Pb+Pb collisions at $\sqrt{s_{NN}}=5.02$
  TeV are shown as dashed lines.}
\label{fig:pt_piKp_LHC}
\end{figure}

We now compare the results of the AMPT+Hydro hybrid 
model calculations 
with various experimental data for Pb+Pb collisions at the LHC, at energies 
$\sqrt{s_{NN}}=2.76$ and $5.02$~TeV . 
Results shown in this section are obtained by 
generating 300 AMPT+Hydro events per centrality bin up to 40\% 
centrality, and 500 events per bin above 40\%.

Figure
\ref{fig:pt_piKp_LHC} shows the transverse momentum spectra of pions,
kaons and protons 
in the $0-5$\% and $30-40$\%
central Pb+Pb collisions at $\sqrt{s_{NN}}=2.76$ TeV in comparison
with the ALICE data at midrapidity \cite{Abelev:2013vea}. The hybrid
model shows a good agreement with the $\pi^+$ and $K^+$ spectra up to
$p_T \sim 2$ GeV. The protons being heavier undergo a strong
blue-shift due to the radial flow. Our results for protons agree quite well
with the data at high $p_T$.  The over-prediction in the proton yields
at low $p_T$, may be due to the neglect of massive hadrons ($m \geq
2.2$ GeV) and final-state hadron rescattering. Also shown are the
predictions of identified hadron spectra for Pb+Pb collisions at
$\sqrt{s_{NN}}=5.02$ TeV (dashed lines). The larger initial temperature 
at this higher collision energy leads to somewhat harder particle
spectra~\cite{Noronha-Hostler:2015uye}.
Note that AMPT alone (with string melting) yields $p_T$ spectra which
are too soft~\cite{Lin:2004en}, so that coupling to hydrodynamics
improves agreement with data.

\begin{figure}[t]
 \includegraphics[width=\linewidth]{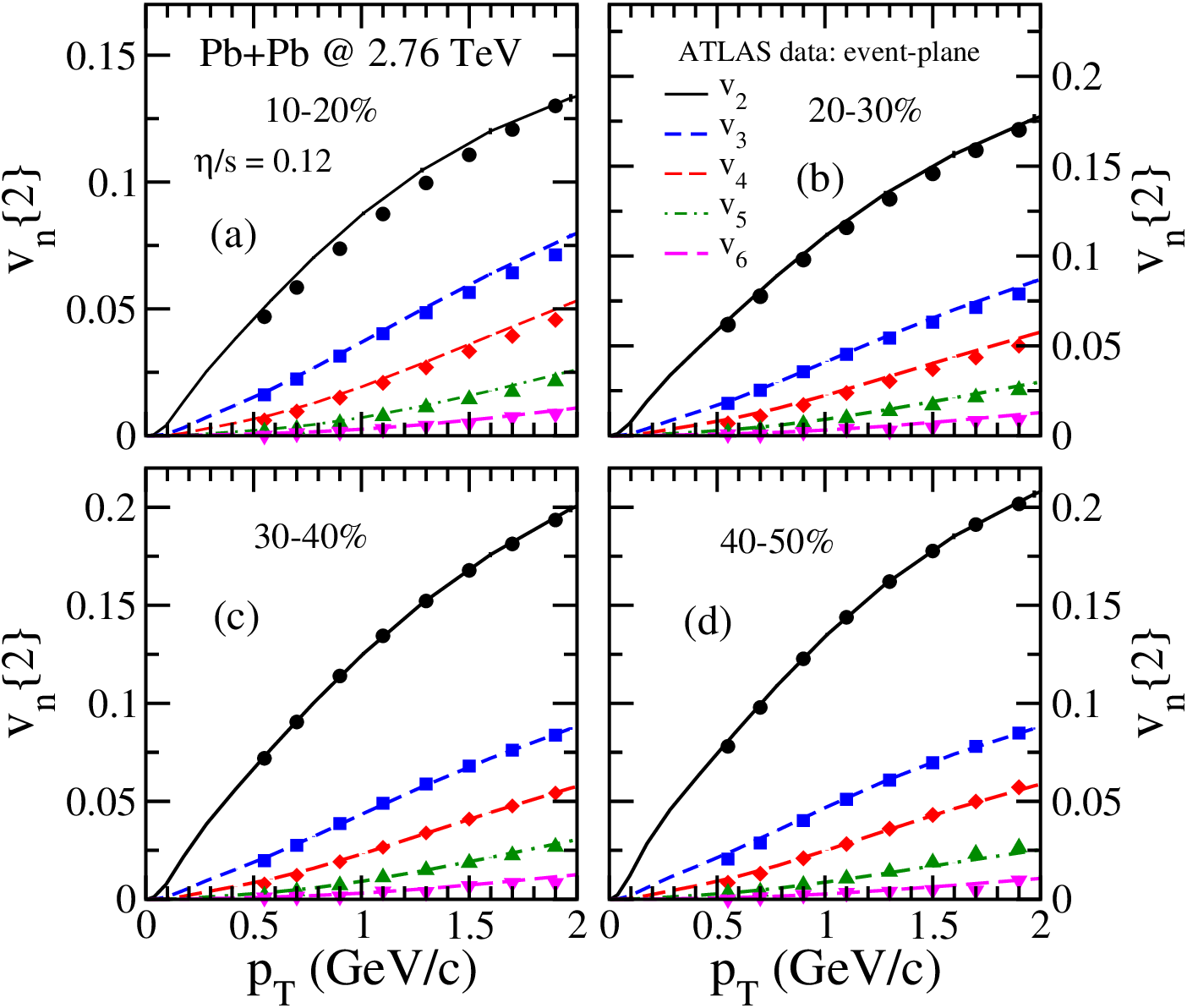}
 \caption{(Color online) Anisotropic flow of charged hadrons 
(top to bottom: $v_2$ to $v_6$) as a function of transverse momentum
in Pb+Pb collisions at   $\sqrt{s_{NN}}=2.76$ TeV in 4 centrality
windows. 
Lines: AMPT+Hydro calculations; 
Symbols: ATLAS data~\cite{ATLAS:2012at}.}
\label{fig:vnpt_LHC276}
\end{figure}
Figure \ref{fig:vnpt_LHC276} compares the anisotropic flow of charged 
hadrons ($v_2$ to $v_6$) from our simulation with the event-plane
results from the 
ATLAS Collaboration \cite{ATLAS:2012at} at $\sqrt{s_{NN}}=2.76$ TeV 
Pb+Pb collisions for various centralities. 
As explained in Sec.~\ref{s:preequilibrium}, our results are obtained by a 
two-particle correlation method, which differs only slightly~\cite{Ollitrault:2009ie}
from the event-plane method used by ATLAS, for realistic values of the event-plane resolution. 
Our hybrid calculations are in good agreement with data over the 
entire $p_T$ range studied, for all the flow harmonics $n=2-6$, and
over a broad centrality range. 

Note that by coupling AMPT to hydrodynamics, we have introduced two free parameters, 
the width $\sigma$ in Eq.~(\ref{smear}) and the viscosity over entropy ratio 
$\eta/s$ (recall that results are essentially independent of the switching time $\tau_{\rm sw}$ 
if one keeps the full $T^{\mu\nu}$ when switching from AMPT to hydrodynamics, 
as shown in Sec.~\ref{s:preequilibrium}). 
Larger $\eta/s$ reduces $v_n$~\cite{Romatschke:2007mq}, 
and smaller $\sigma$ increases the granularity and increases $v_3$. 
The chosen values $\sigma=0.8$~fm and $\eta/s=0.12$ optimize the description of LHC data. 

In particular, agreement is better for this AMPT+hydro model than with
AMPT alone, which underpredicts $v_n$ already at $p_T=2$~GeV~\cite{Han:2011iy,Bzdak:2014dia}.
Note that most initial-state models with subsequent hydrodynamic evolution are
found incompatible with all the flow harmonics even at a given
collision centrality~\cite{Naboka:2015qra}

\begin{figure}[t]
\includegraphics[width=\linewidth]{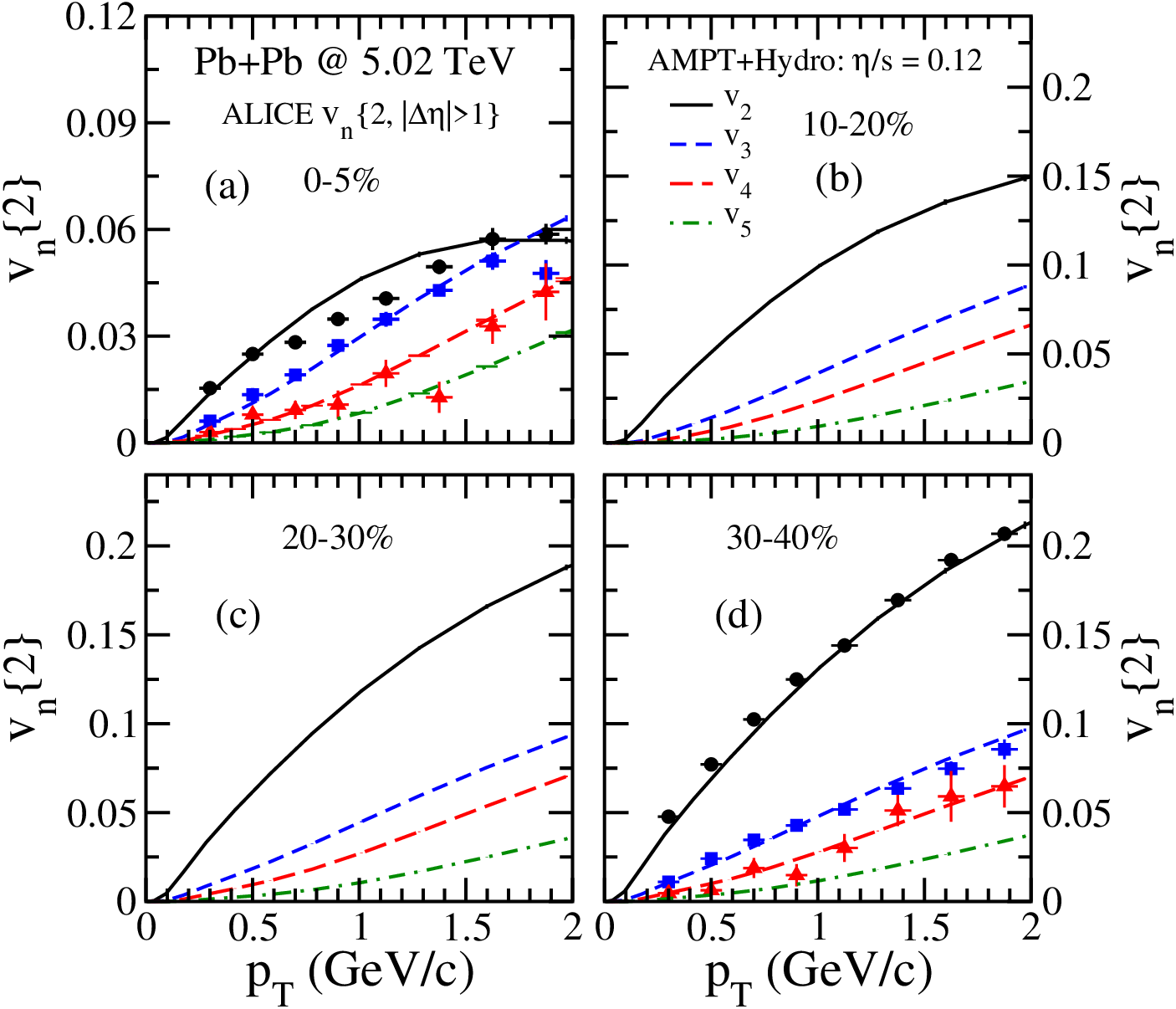}
\caption{(Color online) 
Anisotropic flow in Pb+Pb collisions at $\sqrt{s_{NN}}=5.02$~TeV 
in 4 centrality windows. 
Lines: AMPT+Hydro calculations; 
Symbols: ALICE data for $v_2$, $v_3$ and $v_4$~\cite{Adam:2016izf}.}
\label{fig:vnpt_LHC502}
\end{figure}

The study is extended to the higher energy $\sqrt{s_{NN}}=5.02$~TeV in
Fig.~\ref{fig:vnpt_LHC502}. Panels (a) and (d) display ALICE 
data~\cite{Adam:2016izf}. 
The higher collision energy ensures a slightly larger $v_n(p_T)$
as the VISH2+1 starts with a somewhat higher initial flow anisotropy.
Further, the stronger radial flow blue-shifts the anisotropies to
higher $p_T$, especially for the heavier charged hadrons
\cite{Noronha-Hostler:2015uye,McDonald:2016vlt}. The model provides a
good description of the $v_n(p_T)$, (n=2 to 4) data at $30-40\%$
centrality, and $v_n(p_T)$, (n=3 to 4) data at $0-5\%$ centrality.
It, however, over-predicts somewhat the $v_2$ data at intermediate 
$p_T$ for the $0-5\%$ centrality collisions. Panels (b) and (c) in
Fig. \ref{fig:vnpt_LHC502} present our predictions at two other
centralities.

\begin{figure*}[t]
\begin{center}
\scalebox{0.48}{\includegraphics{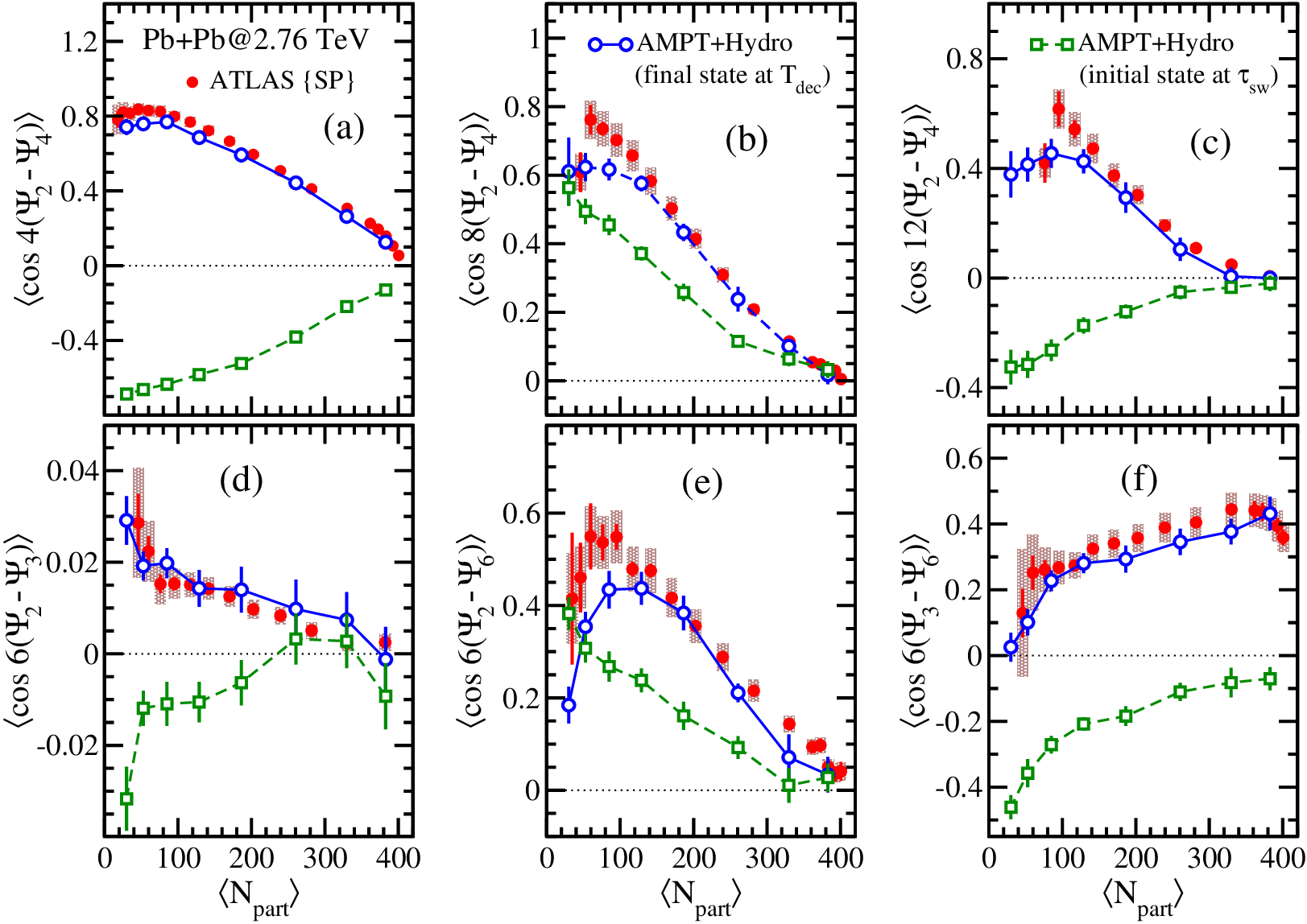}}
\end{center}
\vspace{-0.4cm}
 \caption{(Color online) Two-plane correlations obtained in the initial-state (open squares) and
   final-state (open circles) as a function of the number of
   participants in Pb+Pb collisions at $\sqrt{s_{NN}} = 2.76$ TeV in
   the AMPT+Hydro model as compared to the ATLAS
   data~\cite{Jia:2012sa} using the scalar-product method (solid circles).}
\label{fig:2p}
\end{figure*}

\begin{figure}[b]
 \begin{center}
 \scalebox{0.44}{\includegraphics{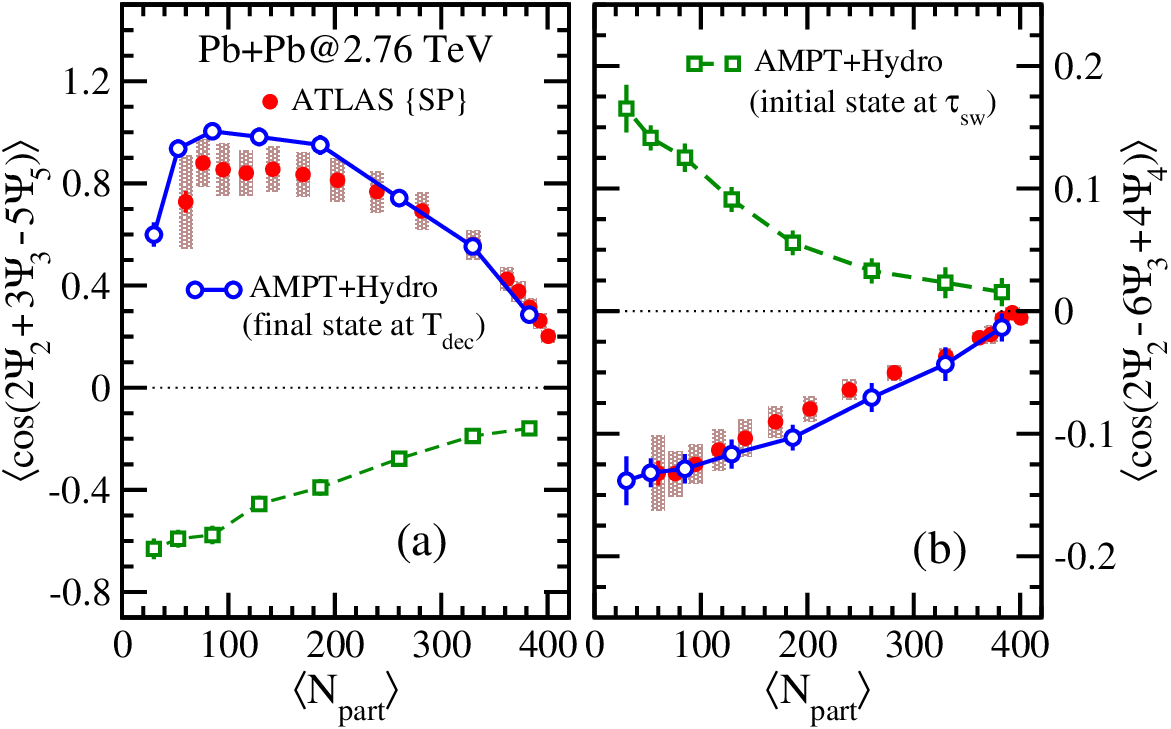}}
 \end{center}
 \vspace{-0.4cm}
 \caption{(Color online) Same as Fig. \ref{fig:2p}, but for
   three-plane correlations.}
 \label{fig:3p}
\end{figure}

Correlations between event planes $\Psi_n$ of different harmonics 
represent higher-order
correlations which can provide crucial information on the
initial-state of the matter~\cite{ALICE:2011ab,Jia:2012sa}
 and on the hydrodynamic response~\cite{Teaney:2013dta}.  The
 ATLAS~\cite{Jia:2012sa}  and ALICE~\cite{Acharya:2017zfg} Collaborations 
have measured several such correlations
between different harmonics $\Psi_n$ and $\Psi_m$ (with $n\neq m$). 
There are two-plane correlations, such as: 
\begin{equation}\label{2pl_sp}
\langle \cos \: 4 (\Psi_2-\Psi_4)\rangle_w
\equiv \frac{\avgev{V_2^2 V^*_4}}
{\sqrt{\avgev{V_2^2 V_2^{*2}}}\sqrt{\avgev{V_4 V^*_4}}},
\end{equation}
where the left-hand side is the quantity measured by ATLAS using the scalar-product method~\cite{Luzum:2012da,Bhalerao:2013ina}, 
and the
right-hand side its expression in a hydrodynamic 
calculation~\cite{Teaney:2013dta}, where $V_n$ is the complex anisotropic
flow, and angular brackets represent an average over 
events in a centrality class. 
Similarly,
the three-plane correlator between harmonics
2, 3, and 5 is defined by ATLAS as\footnote{ALICE uses a slightly different 
normalization~\cite{Acharya:2017zfg,Yan:2015jma}.}
\begin{equation}\label{3pl_sp}
\langle \cos \:  2\Psi_2+3\Psi_3-5\Psi_5)\rangle_w
\equiv \frac{\avgev{V_2 V_3 V_5^*}}
{\sqrt{ \avgev{V_2 V_2^*} \avgev{V_3 V_3^*} \avgev{V_5 V_5^*}}}.
\end{equation}
The two-plane and three-plane correlators
evaluated in this paper are listed in Table I of Ref.~\cite{Bhalerao:2013ina}.
In our calculation of event-plane correlations, we
use the same cuts as ATLAS \cite{Jia:2012sa},
viz. $\eta_c \equiv |\eta| = 0-2.5$ and $p_{T_{\rm min}} = 0.5$ GeV.

Figure \ref{fig:2p} displays the centrality dependence of two-plane
correlations in our AMPT+Hydro calculations. Theoretical results are 
in good agreement with ATLAS data.  
Most correlations are large, and driven by the 
noinlinear hydrodynamic response that couples $v_4$ to $(v_2)^2$ and $v_6$ to $(v_2)^3$
\cite{Yan:2015lwn}. 
Their increase from central to
peripheral collisions is dominated by the increase of $v_2$.
The only exception is the  correlation between 
$\Psi_2$ and $\Psi_3$ (panel (d)), which is much smaller and whose interpretation 
in terms of hydrodynamic response is less simple~\cite{Teaney:2013dta}. 
This correlation is also very well described by our event-by-event calculation.  

Also shown here are the initial-state correlations calculated with the
participant-plane angles $\Phi_n$~\cite{Teaney:2010vd}: 
\begin{equation}\label{eccen}
\varepsilon_n e^{in\Phi_n} \equiv - \frac{\int d^2r_\perp \: \gamma({\bf r}_\perp) 
\: e({\bf r}_\perp) \: r^n_\perp \: e^{in\phi}}
{\int d^2r_\perp \: \gamma({\bf r}_\perp) \: e({\bf r}_\perp) \: r^n_\perp},
\end{equation}
where $e({\bf r}_\perp)$ is the initial energy density, $\gamma({\bf r}_\perp)$ 
is the Lorentz contraction factor due to the transverse flow~\cite{Liu:2015nwa}, 
the integral runs over the transverse plane in a centered coordinate 
system~\cite{Teaney:2010vd}. 
These correlations characterize the initial stage of the hydrodynamic calculations.  
Our results are qualitatively consistent with those presented in~\cite{Qiu:2012uy}.

The centrality dependence of the three-plane correlations is shown in
Fig. \ref{fig:3p}. Here again, the final-state correlations are in good
agreement with the ATLAS data \cite{Jia:2012sa}.
The correlation between $\Psi_2$, $\Psi_3$ and $\Psi_5$ (Fig. \ref{fig:3p} (a)) is large 
and driven by the nonlinear response. That between $\Psi_2$, $\Psi_3$ and $\Psi_4$
(Fig. \ref{fig:3p} (b)), on the other hand, is smaller in magnitude and lacks a simple 
explanation in terms of nonlinear response~\cite{Teaney:2013dta}, but is well 
reproduced in event-by-event hydrodynamics~\cite{Qiu:2012uy}.

\begin{figure}[t]
 \begin{center}
 \scalebox{0.47}{\includegraphics{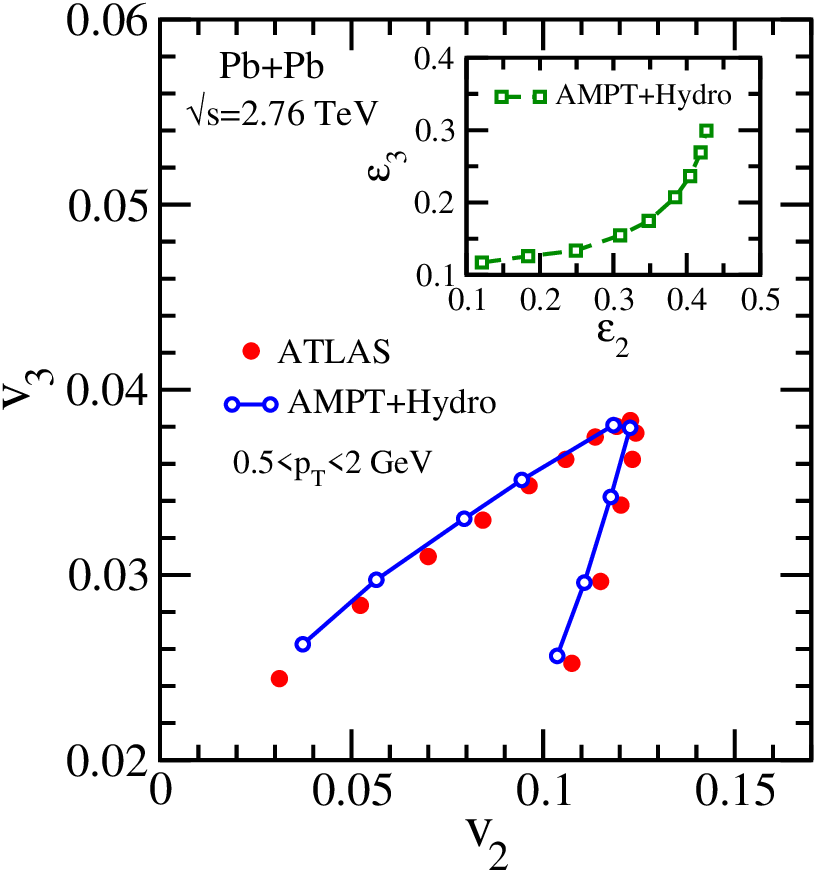}}
 \end{center}
 \vspace{-0.4cm}
 \caption{(Color online) The correlation between $v_2$ and $v_3$ for
   $0.5 < p_T < 2$ GeV/$c$ in Pb+Pb collisions at $\sqrt{s_{NN}} =
   2.76$ TeV in the AMPT+Hydro model (blue open circles) with $\eta/s
   = 0.12$ as compared to the ATLAS data~\cite{Aad:2015lwa} (red solid
   circles). The data points (starting at bottom left) correspond to
   fourteen 5\% centrality intervals over the centrality range 0-70\%.
   The inset shows $\varepsilon_2$-$\varepsilon_3$ correlation as a function
   of centrality in the model calculations.}
\label{fig:v2v3}
\end{figure}

Figure 
\ref{fig:v2v3} displays the centrality dependence of $v_2$ and $v_3$ in the 
($v_2,v_3$) plane,  measured by  ATLAS \cite{Aad:2015lwa}
in Pb+Pb collisions at $\sqrt{s_{NN}} =2.76$ TeV, together with our 
calculation in AMPT+Hydro.  
A boomerang-like shape is observed. 
The corresponding plot for the  initial eccentricities $\varepsilon_2$ and $\varepsilon_3$,
calculated from Eq. (\ref{eccen}) at the switching time $\tau_{\rm sw} =
0.4$ fm/$c$, is also shown in the inset of Fig. \ref{fig:v2v3}. In most
central collisions $\varepsilon_2 \approx \varepsilon_3$, and $\varepsilon_2$
increases faster than $\varepsilon_3$ up to about $45\%$ centrality. For
more peripheral collisions, the large fluctuations in the small
initial geometry contribute to faster rise in $\varepsilon_3$ than
$\varepsilon_2$. In fact, the turning around seen in the $v_2$-$v_3$ plane,
occurs at centralities around $\sim 40$-45\%, thereafter the harmonics
$v_2$ and $v_3$ both decrease. Here, the conversion of
initial spatial asymmetry to final momentum anisotropy is less
efficient due to short lifetime of the plasma, especially for
$\varepsilon_3$ that originates from small-scale structures
(fluctuations).

\begin{figure}[t]
 \begin{center}
 \scalebox{0.47}{\includegraphics{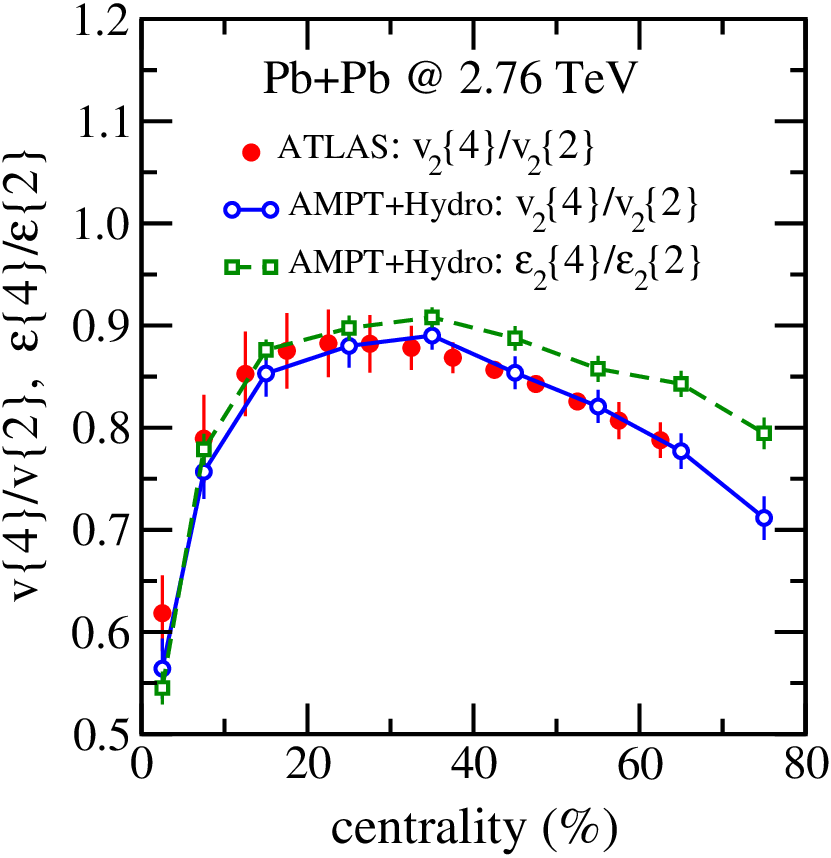}}
 \end{center}
 \vspace{-0.4cm}
 \caption{(Color online) The centrality dependence of the ratio
   $v_2\{4\}/v_2\{2\}$ for the elliptic flow obtained from 2-and
   4-particle cumulant method in Pb+Pb collisions at $\sqrt{s_{NN}} =
   2.76$ TeV in the AMPT+Hydro model (blue open circles) with $\eta/s =
   0.12$ as compared to the ATLAS data~\cite{Aad:2014vba} (red solid
   circles). Also shown is the ratio
   $\varepsilon_2\{4\}/\varepsilon_2\{2\}$ for the eccentricities in the
   model calculations (green open squares).}
\label{fig:ratio_cum}
\end{figure}

Finally, we study event-by-event elliptic flow fluctuations. 
Cumulants~\cite{Borghini:2001vi} of the distribution of $v_2$ differ from one another  
if $v_2$ fluctuates event to event~\cite{Miller:2003kd}. 
The relative fluctuations can be measured through the ratio of the first two cumulants, 
$v_2\{4\}/v_2\{2\}$. 
The fluctuations of $v_2$ originate to a large extent from the fluctuations of the
initial eccentricity $\varepsilon_2$~\cite{Alver:2006wh}. 
If  $v_2$ is proportional to  $\varepsilon_2$, that is, 
if $v_2/\varepsilon_2$ is the same for all events in a centrality class~\cite{Niemi:2012aj}, 
then $v_2\{4\}/v_2\{2\}$ coincides with $\varepsilon_2\{4\}/\varepsilon_2\{2\}$. 
Event-by-event hydrodynamics allows to directly test this relation by computing both quantities. 

In Fig.~\ref{fig:ratio_cum}, we 
compare the centrality dependence of the initial and final cumulant
ratios, $\varepsilon_2\{4\}/\varepsilon_2\{2\}$ and $v_2\{4\}/v_2\{2\}$, 
in Pb+Pb collisions at $\sqrt{s_{NN}} = 2.76$ TeV in the AMPT+hydro model.
The ratios coincide for central collisions, but $v_2\{4\}/v_2\{2\}$ becomes smaller 
than $\varepsilon_2\{4\}/\varepsilon_2\{2\}$ as the centrality percentile increases. 
This trend has already been observed in hydrodynamic calculations~\cite{Giacalone:2017uqx} 
and attributed to a nonlinear (cubic) response~\cite{Noronha-Hostler:2015dbi}. 
Our results are in excellent agreement with ATLAS data~\cite{Aad:2014vba} over the entire centrality range. 
Note that they would not agree if $v_2$ was simply proportional to $\varepsilon_2$ in every event, 
as already observed with a different model of initial conditions~\cite{Giacalone:2017uqx}.
This suggests that the success of hydrodynamics in describing elliptic flow fluctuations extends beyond 
a mere linear response to the initial eccentricity. 

\section{Conclusions}

We have studied the effects of preequilibrium dynamics in heavy-ion collisions 
by modeling the early stages using a transport calculation with realistic cross sections, 
and coupling it to a (2+1)-dimensional viscous hydrodynamic calculation to describe the later evolution. 
Our model of the initial stage describes the microscopic dynamics of quarks and antiquarks 
as soon as they are produced, as modeled in AMPT. 
The initialization of the hydrodynamic calculation takes into account the fact that the transverse 
momenta of partons at a given point do not add up to zero  and that they are not in 
local equilibrium: initial transverse flow and initial shear pressure are thus naturally taken into account. 
We have thus set up a comprehensive framework to perform calculations, 
which couples consistently initial stage dynamics and hydrodynamic evolution. 

We have studied the effects of preequilibrium dynamics by switching off its components one by one. 
Initial transverse flow results in harder momentum spectra and larger anisotropic flow. 
This effect is more pronounced if the switching time from AMPT to hydrodynamics is delayed. 
The initial shear viscous pressure $\pi^{\mu\nu}$  has a much smaller effect: this is explained 
by our observation that various initializations of $\pi^{\mu\nu}$
relax to a common value at an early time
$(\tau-\tau_{\rm sw}) \simeq 1$ fm/$c$ and remain similar in
magnitude thereafter. 
When the full preequilibrium dynamics is taken into account in initializing the hydrodynamic calculation, 
final results are insensitive to the choice of the switching time.

The model, with full initial dynamics ($v_T \neq 0 \neq
\pi^{\mu\nu}$), describes identified particle spectra and differential anisotropic flow $v_n(p_T) ~(n=2-6)$
at various centralities for Pb+Pb collisions at the LHC, with a constant
shear viscosity to entropy density ratio of $\eta/s = 0.12$.
We have also tested our formalism against quantities which had not yet been computed in the AMPT+hydro framework, 
in particular event-plane correlations and elliptic flow fluctuations, which
probe the initial conditions and the hydrodynamic response in an 
independent way. Our calculations for these quantities are also in 
excellent agreement with LHC data.
This overall agreement 
 suggests that the AMPT model provides a reasonable description of the early stages
of nucleus-nucleus collisions, and confirms the usual statement that the quark-gluon plasma 
produced at the LHC has a low shear viscosity over entropy ratio. 

\section*{Acknowledgments}

RSB would like to acknowledge the hospitality of the IPhT, Saclay,
France where a part of this work was done and the support of the CNRS
LIA (Laboratoire International Associ\'{e}) THEP (Theoretical High
Energy Physics) and the INFRE-HEPNET (IndoFrench Network on High
Energy Physics) of CEFIPRA/IFCPAR (Indo-French Center for the
Promotion of Advanced Research). RSB also acknowledges the support of
the Department of Atomic Energy, India for the award of the Raja Ramanna
Fellowship.
JYO thanks Giuliano Giacalone and Bj\"orn Schenke for useful comments
on the manuscript.

\end{document}